\documentclass[onecolumn,showpacs,preprintnumbers,prc,floatfix]{revtex4}
\usepackage{amsmath,amssymb}
\usepackage{graphicx}
\usepackage{dcolumn}
\usepackage[vcentermath]{youngtab}
\usepackage[dvips]{epsfig,color}
\usepackage{bm}
\begin{document}
\title{Strange and nonstrange baryon spectra in the interacting quark-diquark model\footnote{Talk presented at the NSTAR 2015 Conference, May 25-28 2015, Osaka (Japan).}}   
\author{J. Ferretti}
\affiliation{Dipartimento di Fisica and INFN, Universit\`a di Roma "Sapienza", Piazzale A. Moro 5, I-00185 Roma, Italy}
\begin{abstract}
We briefly discuss the relativistic interacting quark-diquark model formalism and its application to the calculation of strange and nonstrange baryon spectra. The results are compared to the existing experimental data. 
\end{abstract}
\maketitle

\section{Introduction}
In quark-diquark models, baryons are assumed to be composed of a constituent quark, $q$, and a constituent diquark, $Q^2$ \cite{Ida:1966ev,lich}.
The effective degree of freedom of diquark, introduced by Gell-Mann in his original paper on quarks \cite{GellMann:1964nj}, has been used in a large number of studies, from one-gluon exchange to lattice QCD calculations \cite{Jakob:1997,Brodsky:2002,Gamberg:2003,Jaffe:2003,Wilczek:2004im,Jaffe:2004ph,Selem:2006nd,DeGrand:2007vu,BacchettaRadici,Forkel:2008un,Anisovich:2010wx} and, more recently, also in tetraquark spectroscopy \cite{Maiani:2004vq,Santopinto:2006my}. 

Up to an energy of 2 GeV, the diquark can be described as two correlated quarks with no internal spatial excitations \cite{Santopinto:2004hw,Ferretti:2011zz}. 
Then, its color-spin-flavor wave function must be antisymmetric. 
Moreover, as we consider here only light baryons, made up of $u$, $d$, $s$ quarks, the internal group is restricted to SU$_{\mbox{sf}}$(6). 
If we denote spin by its value, flavor and color by the dimension of the representation, the quark has spin $s_2 = \frac{1}{2}$, flavor $F_2={\bf {3}}$, and color $C_2 = {\bf {3}}$. 
The diquark must transform as ${\bf {\overline{3}}}$ under SU$_{\mbox{c}}$(3), hadrons being color singlets. Then, one only has the symmetric SU$_{\mbox{sf}}$(6) representation $\mbox{{\boldmath{$21$}}}_{\mbox{sf}}$(S), containing $s_1=0$, $F_1={\bf {\overline{3}}}$, and $s_1=1$, $F_1={\bf {6}}$, i.e., the scalar and axial-vector diquarks, respectively \cite{Wilczek:2004im,Jaffe:2004ph}. If we indicate the possible diquark states by their constituent quark content in square (scalar diquarks) or brace brackets (axial-vector diquarks), then the possible scalar diquark configurations are $[n,n]$ and $[n,s]$ (where $s$ is a strange quark, while $n = u,d$) , while the possible axial-vector diquark configurations are $\{n,n\}$, $\{n,s\}$ and $\{s,s\}$ \cite{Jaffe:2004ph}.

In this contribution, we discuss the relativistic interacting quark-diquark model of Refs. \cite{Santopinto:2004hw,Ferretti:2011zz,DeSanctis:2011zz,qD2014a,qD2014b}, which is a potential model for strange and nonstrange baryon spectroscopy constructed within the point form formalism \cite{Klink:1998zz,Pavia-Graz,Sanctis:2007zz}.
In our model, baryon resonances are described as two-body quark-diquark bound states, thus the relative motion between the two constituents and the Hamiltonian of the model are functions of the relative coordinate $\vec r$ and its conjugate momentum $\vec q$. 
The Hamiltonian contains a Coulomb plus linear confining interaction and an exchange one, depending on the spins and isospins of the quark and the diquark.
The strange and nonstrange spectra are computed and the results compared to the existing experimental data \cite{Nakamura:2010zzi}.

\section{The Mass operator}
\label{The Model} 
We consider a quark-diquark system, where $\vec{r}$ and $\vec{q}$ are the relative coordinate between the two constituents and its conjugate momentum, respectively. 
The baryon rest frame mass operator we consider is 
\begin{equation}
	\begin{array}{rcl}
	M & = & E_0 + \sqrt{\vec q\hspace{0.08cm}^2 + m_1^2} + \sqrt{\vec q\hspace{0.08cm}^2 + m_2^2} 
	+ M_{\mbox{dir}}(r)  \\
	& + & M_{\mbox{ex}}(r)  
	\end{array}  \mbox{ },
	\label{eqn:H0}
\end{equation}
where $E_0$ is a constant, $M_{\mbox{dir}}(r)$ and $M_{\mbox{ex}}(r)$ respectively the direct and the exchange diquark-quark interaction, $m_1$ and $m_2$ stand for diquark and quark masses, where $m_1$ is either $m_{[q,q]}$ or $m_{\{q,q\}}$ according if the mass operator acts on a scalar or axial-vector diquark.
The direct term we consider, 
\begin{equation}
  \label{eq:Vdir}
  M_{\mbox{dir}}(r)=-\frac{\tau}{r} \left(1 - e^{-\mu r}\right)+ \beta r ~~,
\end{equation}
is the sum of a Coulomb-like interaction with a cut off plus a linear confinement term. 
We also have an exchange interaction, since this is the crucial ingredient of a quark-diquark description of baryons \cite{Santopinto:2004hw,Lichtenberg:1981pp}. 
In the nonstrange sector we have \cite{Santopinto:2004hw,Ferretti:2011zz}
\begin{equation}
	\begin{array}{rcl}
	M_{\mbox{ex}}(r) & = & \left(-1 \right)^{L + 1} \mbox{ } e^{-\sigma r} \left[ A_S \mbox{ } \vec{s}_1 
	\cdot \vec{s}_2  \right. \\ 
	& + & \left. A_I \mbox{ } \vec{t}_1 \cdot \vec{t}_2  
	+  A_{SI} \mbox{ } \vec{s}_1 \cdot \vec{s}_2 \mbox{ } \vec{t}_1 \cdot \vec{t}_2  \right]  
	\end{array}  \mbox{ },
	\label{eqn:Vexch-nonstrange}
\end{equation}
where $\vec{s}$ and $\vec{t}$ are the spin and the isospin operators, while for strange baryons we consider a G\"ursey-Radicati inspired interaction \cite{Gursey:1992dc,qD2014b}.
In the nonstrange sector, we also have a contact interaction 
\begin{equation}
	\begin{array}{rcl}
	\label{eqn:Vcont}	
	M_{\mbox{cont}} & = & \left(\frac{m_1 m_2}{E_1 E_2}\right)^{1/2+\epsilon} \frac{\eta^3 D}{\pi^{3/2}} 
	e^{-\eta^2 r^2} \mbox{ } \delta_{L,0} \delta_{s_1,1} \\ 
	& \times & \left(\frac{m_1 m_2}{E_1 E_2}\right)^{1/2+\epsilon}
	\end{array}  \mbox{ },
\end{equation}
introduced in the mass operator of Ref. \cite{Ferretti:2011zz} to reproduce the $\Delta-N$ mass splitting.

\section{Results and discussion}
\begin{figure}[htbp]
\begin{center}
\includegraphics[width=7cm]{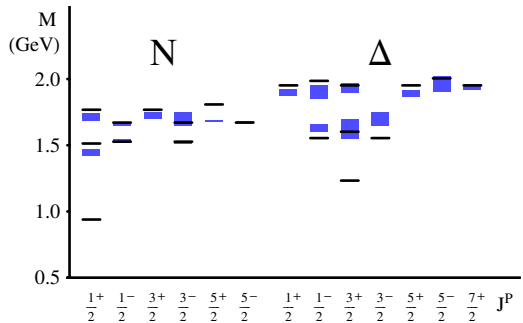}
\end{center}
\caption{Comparison between the calculated masses (black lines) of the $3^*$ and $4^*$ $N$ and $\Delta$ resonances (up to 2 GeV) and the experimental masses from PDG \cite{Nakamura:2010zzi} (blue boxes).} 
\label{fig:Spectrum-ND}
\end{figure}

In this section, we show our results for the non-strange baryon spectrum from Ref. \cite{Ferretti:2011zz}. See Fig. \ref{fig:Spectrum-ND}. 

While the values of the diquarks masses $m_n$, $m_{[n,n]}$ and $m_{\{n,n\}}$ almost coincide in the "strange" and "nonstrange" fits \cite{Ferretti:2011zz,qD2014b}, there is a certain difference between the values of a few model parameters used in the two fits. This is especially evident in the case of the exchange potential parameters, $A_S$ and $A_I$. This difference is due to the substitution of the spin-isospin term in the exchange potential with the $SU(3)$ flavor-dependent, which also determines a change in the values of the spin and isospin, $A_S$ and $A_I$, parameters.
Moreover, and most important, some parameters are present in one fit and not in the other, such as the contact interaction ones, because the potential of Eq. (\ref{eqn:Vcont}) was introduced to reproduce the $\Delta-N$ mass splitting, and thus it is inessential in the strange sector. 
It is also interesting to note that in our model $\Lambda(1116)$ and $\Lambda^*(1520)$ are described as bound states of a scalar diquark $[n,n]$ and a quark $s$, where the quark-diquark system is in $S$ or $P$-wave, respectively \cite{qD2014b}. This is in accordance with the observations of Refs. \cite{Jaffe:2004ph,Selem:2006nd} on $\Lambda$'s fragmentation functions, that the two resonances can be described as $[n,n]-s$ systems. 

%%%%%%%%%%%%%%%%%%%%%%%%%%%%%%%%%%%%%%%%
%%%%%%%%%%%%%%%%%%%%%%%%%%%%%%%%%%%%%%%%

It is interesting to compare the present results to those of the main three-quark quark models \cite{IK,CI,HC,GR,LMP}. It is clear that a larger number of experiments and analyses, looking for missing resonances, are necessary because many aspects of hadron spectroscopy are still unclear \cite{Hugo}.

The present work can be expanded to include charmed and/or bottomed baryons \cite{FS-inprep}, which can be quite interesting in light of the recent experimental effort to study the properties of heavy hadrons.

%%%%%%%%%%%%%%%%%%%%%%%%%%%%%%%%%%%%%%%%%%%%%%%%%%

\end{document}